\documentclass[sigconf,10pt,screen]{acmart}

\copyrightyear{2025}
\acmYear{2025}
\setcopyright{cc}
\setcctype{by}
\acmConference[ICS '25]{2025 International Conference on Supercomputing}{June 8--11, 2025}{Salt Lake City, UT, USA}
\acmBooktitle{2025 International Conference on Supercomputing (ICS '25), June 8--11, 2025, Salt Lake City, UT, USA}
\acmDOI{10.1145/3721145.3730422}
\acmISBN{979-8-4007-1537-2/2025/06}

\setcopyright{acmlicensed}

\usepackage{booktabs}   

\usepackage{algorithmic}
\usepackage{graphicx}
\usepackage{textcomp}
\usepackage{xcolor}

\usepackage{amsthm}

\usepackage[ruled,vlined, linesnumbered]{algorithm2e}

\usepackage{multirow}
\usepackage{todonotes}

\usepackage{pifont}
\usepackage{makecell}

\usepackage{subfigure}
\usepackage{comment}

\usepackage[skip=3pt,font=normal]{caption}
\usepackage{enumitem}
\usepackage{hyperref}
\usepackage{url}
\DeclareRobustCommand*\circled[1]{\tikz[baseline=(char.base)]{
            \node[shape=circle,fill,inner sep=1.3pt] (char) {\textcolor{white}{#1}};}}

\newcommand{\zhuang}[1]{\textcolor{black}{#1}}

\newcommand{\zhuangppopp}[1]{\textcolor{black}{#1}}

\newcommand{\zhuanghpdc}[1]{\textcolor{black}{#1}}
\newcommand{\lingqi}[1]{\textcolor{black}{#1}}
\newcommand{\jiajun}[1]{\textcolor{black}{#1}}

\newcommand{\wu}[1]{\textcolor{black}{#1}}

\newcommand{\method}{\texttt{SuperGCN}}

\def\BibTeX{{\rm B\kern-.05em{\sc i\kern-.025em b}\kern-.08em
    T\kern-.1667em\lower.7ex\hbox{E}\kern-.125emX}}

\begin{document}

\title{Scaling Large-scale GNN Training to Thousands of Processors on CPU-based Supercomputers
}

\newcommand{\affsym}[1]{$^{#1}$}

\author{Chen Zhuang\affsym{1,2}, Lingqi Zhang\affsym{2}, Du Wu\affsym{1,2}, Peng Chen\affsym{2}$^*$, Jiajun Huang\affsym{3}, Xin Liu\affsym{4}, Rio Yokota\affsym{1}, Nikoli Dryden\affsym{5}, Toshio Endo\affsym{1}, Satoshi Matsuoka\affsym{2,1}, Mohamed Wahib\affsym{2}$^*$\thanks{$^*$ Corresponding authors}}

\affiliation{
    \institution{\affsym{1}Institute of Science Tokyo, Japan, \affsym{2}RIKEN Center for Computational Science, Japan,}
    \institution{\affsym{3}University of South Florida, USA, \affsym{4}National Institute of Advanced Industrial Science and Technology, Japan, \affsym{5}Lawrence Livermore National Laboratory, USA}
    \institution{}
    \country{}
}

\email{{chen.zhuang, lingqi.zhang, du.wu, peng.chen, mohamed.attia}@riken.jp, jiajun.huang.cs@gmail.com,}
\email{xin.liu@aist.go.jp, rioyokota@rio.scrc.iir.isct.ac.jp, dryden1@llnl.gov,endo@scrc.iir.isct.ac.jp, matsu@acm.org}


\renewcommand{\shortauthors}{Zhuang et al.}
\renewcommand{\authors}{Chen Zhuang, Lingqi Zhang, Du Wu, Peng Chen, Jiajun Huang, Xin Liu, Rio Yokota, Nikoli Dryden, Toshio Endo, Satoshi Matsuoka and Mohamed Wahib}


\begin{abstract}

Graph Convolutional Networks (GCNs), particularly for large-scale graphs, are crucial across numerous domains. However, training distributed full-batch GCNs on large-scale graphs suffers from inefficient memory access patterns and high communication overhead. To address these challenges, we introduce \method{}, an efficient and scalable distributed GCN training framework tailored for CPU-powered supercomputers. Our contributions are threefold: (1) we develop general and efficient aggregation operators designed for irregular memory access, (2) we propose a hierarchical aggregation scheme that reduces communication costs without altering the graph structure, and (3) we present a communication-aware quantization scheme to enhance performance.
Experimental results demonstrate that \method{} achieves a speedup of up to 6$\times$ compared with the SoTA implementations, and scales to 1000s of HPC-grade CPUs on the largest publicly available datasets, without sacrificing model convergence and accuracy. Moreover, due to the effective strong scaling of \method{}, we outperform SoTA GPU-based and CPU-based distributed full-batch GCN training frameworks, in absolute performance, for large-scale graphs. 

\end{abstract}




\maketitle

\section{Introduction}
\zhuanghpdc{Graph Convolutional Networks (GCNs) have been widely applied across various domains such as social networks analysis~\cite{guo2019attention}, biology~\cite{jiang2020brain, yuan2020gcng, meng2022boosting}, and chemistry~\cite{sun2020drug, ryu2019bayesian}, due to their ability to process graph-structured data. Notably, GCNs play a crucial role in the recent surge of AI for Science. For instance, some recent studies~\cite{pfaff2020learning, belbute2020combining, cao2023efficient, lam2023learning} have focused on transforming unstructured and adaptive mesh-based domains in physics simulations into graphs, enabling the replacement of traditional iterative solvers with GCNs. These approaches achieve comparable accuracy to conventional iterative methods while significantly reducing computational costs.
The widespread presence of such large-scale graphs in real-world applications has driven the need for distributed GCN frameworks capable of efficiently handling large-scale graphs on large-scale computing systems.}



\zhuanghpdc{Handling large-scale graphs in distributed GCN frameworks typically follows one of two types of training strategies: mini-batch and full-batch. The mini-batch method is widely supported in various studies, such as~\cite{hamilton2017inductive, ying2018pinsage, chen2018fastgcn, chiang2019cluster}. In mini-batch training, the original graph is sampled into small subgraphs, which might destroy the original graph structure information during the training process, leading to potential accuracy degradation~\cite{chen2018stochastic, jia2020roc}.
Hence, the distributed full-batch training strategy, if designed to be scalable, is preferred for large-scale graphs, due to its ability to retain the original graph structure information. }

\zhuanghpdc{Distributed full-batch GCN training is similar to model parallelism in conventional Deep Neural Networks (DNNs), it treats the full graph as the entire training dataset, and partitions the graph into multiple subgraphs to assign them to different workers for computation. }\lingqi{After partitioning the graph, }\zhuanghpdc{the distributed full-batch parallelism scheme applied on the graph is broken down to two distinctive components: local computation at the worker level, and remote communication to communicate data between the subgraphs. This scheme presents two challenges for building an efficient and scalable GCNs training system:  
(1) Irregular memory access pattern and load imbalance caused by the sparsity and randomness of graphs. 
(2) Graph partitioning across workers creates numerous edge cuts between subgraphs, leading to significant imbalanced communication between different workers to preserve the original edge connections~\cite{wan2022bns}.}


\zhuanghpdc{
To address these two challenges of irregular memory access and imbalanced communication, clusters and supercomputers equipped with HPC-grade CPUs and high-performance interconnection networks represent a viable option over the more expensive and power-consuming GPU-accelerated systems. 
Therefore, it is extremely important to develop performant and scalable full-batch GCNs training solutions that are generic for different CPU platforms.
}

\zhuanghpdc{
Currently, there are many existing distributed GCN training systems. For example, regarding the challenge of communication overhead, \cite{wan2022bns} reduces communication volume by sampling communication data, ~\cite{md2021distgnn, wan2022pipegcn, peng2022sancus, thorpe2021dorylus, zhang2024sylvie} use asynchronous communication to overlap with computation, and \cite{wan2023adaptive, zhang2024sylvie} employ adaptive quantization to minimize communication costs. However, these approaches face several limitations: (1) they are not optimized for different CPU platforms, (2) \lingqi{sampling} significantly modifies the graph structure, and leads to accuracy degradation similar to mini-batch training, (3) asynchronous communication breaks the dependency between epochs, introducing known issues such as data staleness and slower model convergence~\cite{dai2018toward}, (4) most importantly, \lingqi{none of the above approaches} have been validated for extremely large graphs \lingqi{using distributed} systems with a large number of machines.
}


In this paper, we \lingqi{design} an efficient, scalable, and general distributed GCN training \lingqi{system} for large-scale graphs on CPU-based supercomputers. The contributions of this paper are as follows:

{\color{black}
\setlength{\leftmargini}{10 pt}
\begin{itemize} 
    \item \textbf{General and Efficient Aggregation Operators.} We propose efficient CPU-level aggregation operators designed to support different CPU platforms (Arm and x86).

    \item \textbf{Hierarchical Aggregation Scheme:} We propose a novel hybrid pre-post aggregation method based on the minimum vertex cover algorithm, reducing communication volume while preserving the original graph structure.

    \item \textbf{Communication-Aware Quantization Scheme:} We leverage aggressive quantization communication to further reduce communication overhead. We additionally integrate masked label propagation and normalization to maintain model accuracy.
    


    \item Experimental results demonstrate that \method{} achieves up to 6$\times$ speedup compared to state-of-the-arts on CPU-based supercomputers, successfully trains on the largest publicly available GNN datasets using thousands of processors, and outperforms SoTA GPU-based and CPU-based distributed full-batch GCN training systems when operating at their respective maximum achievable performance.

\end{itemize}
}

\section{Background}
\label{sec:background}

\begin{figure}[t!]
    \centering
    \includegraphics[clip,width=0.44\textwidth]{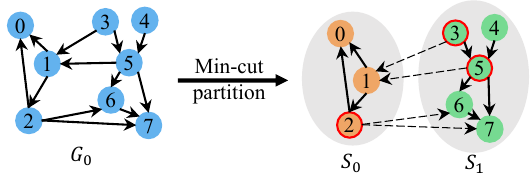}
    \caption{Illustration of graph partitioning. Dotted lines indicate cut edges separating distinct subgraphs. Features of boundary nodes (red circles) need to be transferred across these cut edges via remote communication.}
    \label{fig/background:distributed_gcn}
\end{figure}

 \begin{table*}[!ht]  
      \caption{Comparison of different distributed full-batch GNNs training solutions. }
      \small
      \centering
      \vspace{-4pt}
      \setlength{\tabcolsep}{10pt}
      \resizebox{0.9\linewidth}{!}{
          \begin{tabular}{| l|c|c|c|c|c|c| }
          \hline
          \bf{Methods} & \makecell[c]{\bf{Target} \\ \bf{Platform}} & \makecell[c]{\bf{Optimized} \\ \bf{operators}} & \makecell[c]{\bf{Preserve} \\ \bf{graph structure}} & \makecell[c]{\bf{Fresh} \\ \bf{boundary nodes}} & \makecell[c]{\zhuang{\bf{Reduce comm}} \\ \bf{volume}} & \makecell[c]{\bf{Reported max} \\ \bf{workers number}} \\
          \hline
          \textsf{ROC} \cite{jia2020roc} & GPU & \ding{55} & \ding{52} & \ding{52} (Synchronous) & \zhuang{\ding{55}} & 16 \\
          \textsf{CAGNET} \cite{tripathy2020reducing} & GPU & \ding{55} & \ding{52} & \ding{52} (Synchronous) & \ding{52} & 100 \\
          \textsf{BNS-GCN} \cite{wan2022bns} & GPU & \ding{55} & \ding{55} & \ding{52} (Synchronous) & \ding{52} & 192 \\ 
          \textsf{PipeGCN} \cite{wan2022pipegcn} & GPU & \ding{55} & \ding{52} & \ding{55} (Asynchronous) & \ding{55} & 32 \\
          \textsf{Sancus} \cite{peng2022sancus} & GPU & \ding{55} & \ding{52} & \ding{55} (Asynchronous) & \ding{55} & 8 \\
          \textsf{Dorylus} \cite{thorpe2021dorylus} & Serverless & \ding{55} & \ding{52} & \ding{55} (Asynchronous) & \ding{55} & 32 \\
          \textsf{DGCL} \cite{cai2021dgcl} & GPU & \ding{55} & \ding{52} & \ding{52} (Synchronous) & \ding{55} & 8 \\ 
          \textsf{AdapQ} \cite{wan2023adaptive} & GPU & \ding{55} & \ding{52} & \ding{52} (Synchronous) & \ding{52} & 8 \\ 
          \textsf{SAR} \cite{mostafa2022sequential} & x86 only & \ding{55} & \ding{52} & \ding{52} (Synchronous) & \ding{55} & 128 \\ 
         \textsf{DistGNN} \cite{md2021distgnn} & x86 only & \ding{52} & \ding{52} & \ding{55} (Asynchronous) & \ding{55} & 256 \\
          \bf{\textsf{SuperGCN (ours)}} & \bf{x86 \& Arm} & \bf{\ding{52}} & \bf{\ding{52}} & \bf{\ding{52} (Synchronous)} & \bf{\ding{52}} & \bf{8,192} \\
          \hline
           \end{tabular}  
           }
       \label{table/related_work:related_work} 
\end{table*}     

\subsection{Graph Convolutional Networks}
GCNs consist of multiple graph convolutional layers. GCNs take as input a graph and a feature matrix that stores the features of each node on the graph. The forward pass of GCN layers involves two stages: \emph{Aggregation of neighbors' features} and \emph{NN operations}. 
\zhuangppopp{The stage of \emph{Aggregation of neighbors' features}, represented as {\scriptsize$z^{(l-1)}_{i} = AGGREGATE \left(\left\{h_{j}^{(l-1) } \mid j \in N(i) \right\}\right)$}, involves each node $i$ on the graph aggregates the features of neighboring nodes $N(i)$ using an $AGGREGATE$ function (sum, mean, etc.). Following this is a neural network operation $UPDATE$, such as a linear transform, performed on node $i$'s $l-1^{th}$ layer features $h^{(l-1)}_{i}$, and aggregation result $z^{(l-1)}_{i}$ to obtain the $l^{th}$ layer features $h^{(l)}_{i}$. This stage is referred to as \emph{NN operations},  
{\scriptsize $h^{(l)}_{i} = UPDATE \left(z^{(l-1)}_{i}, h^{(l-1)}_{i}\right)$}.
}
There are several GCN-related models, such as GCN~\cite{kipf2016semi} and GraphSAGE~\cite{hamilton2017inductive}. 
    

\subsection{Distributed Full-batch GCNs Training}

Distributed full-batch GCN training is crucial for addressing memory limitations and reducing training time. The input graph is divided into smaller subgraphs, each managed by a worker, similar to parallelism in traditional distributed neural networks~\cite{DBLP:conf/hpdc/KahiraNBTBW21}. 
Graph partitioning creates "cut edges" (dotted lines in Fig.~\ref{fig/background:distributed_gcn}), where the source and destination nodes belong to different subgraphs. These source nodes, known as \emph{boundary nodes} (highlighted as red circles \emph{2,3,5} in Fig.~\ref{fig/background:distributed_gcn}), require remote communication along cut edges to be transferred to the destination worker in aggregation of GCN layers.
Several works have optimized distributed full-batch GCN training (Table~\ref{table/related_work:related_work}). 
However, few studies target large-scale general CPU supercomputers (x86, Arm) and scale to thousands of processors. 

\subsection{Full-batch vs. Mini-batch GCNs}
\zhuangppopp{Mini-batch GCNs training samples nodes from the original graph to form smaller batches, on which training is performed~\cite{hamilton2017inductive, chiang2019cluster}.
In contrast, full-batch GCNs training uses the entire graph as the training data.} While Mini-batch training reduces memory usage and computation, it can result in loss of graph information. Studies~\cite{hamilton2017inductive, chen2018fastgcn, chiang2019cluster} suggest that sampling-based methods do not compromise accuracy on certain benchmarks, but there is no theoretical proof of their effectiveness across different graphs. The influence of graph properties on sampling in mini-batch training is still not well understood. For instance, Chen et al.~\cite{chen2017stochastic} highlight that the popular neighbor sampling method, \emph{GraphSAGE}, lacks convergence guarantees due to biased predictions. Other studies~\cite{jia2020roc, thorpe2021dorylus} show that mini-batch GCNs perform worse on the Reddit dataset~\cite{hamilton2017inductive}. Moreover, traditional sampling methods~\cite{wang2011understanding} often fail to retain key graph structures, causing information loss in large-scale graphs.


\subsection{Stochastic Integer Quantization}
Stochastic integer quantization~\cite{chen2021actnn} is a fundamental technique in the field of machine learning. It is pivotal in optimizing various aspects of neural network training and communication efficiency~\cite{zhu2020towards, feng2020sgquant, wan2023adaptive}. This technique involves representing real-valued numbers as integers or real-valued numbers with reduced bit widths, thereby reducing memory footprint and accelerating computation. Quantization and dequantization methods can be written as {\footnotesize$h_{quant} = round\{(h - Z)/S\}$} and {\footnotesize $h_{dequant} = h_{quant} * S + Z$}, where {\footnotesize$Z = min(h)$},  {\footnotesize$S = {(max(h) - min(h))}/{(2^b - 1)}$}, and $b$ is the bit width. The $round$ function is the stochastic rounding operation~\cite{jacob2018quantization}.



\subsection{GNNs with Masked Label Propagation}
\zhuanghpdc{Label Propagation (LP) in Graph Neural Networks (GNNs) is a mechanism for distributing label information throughout a graph to assist in learning and prediction tasks. It improves the influence of nodes with the same label in the graph and improves model accuracy~\cite{wang2020unifying}. A specific type of label propagation in GNNs is Masked Label Propagation~\cite{shi2020masked}. In GNN training with Masked Label Propagation, a portion of the labeled nodes' labels is randomly masked during each training step. The remaining unmasked labels are embedded into a vector that has the same dimension as the original feature vectors of nodes and added (sum) to the original feature vectors of nodes. With this step, in the message-passing operation (aggregation) of GNNs where the feature vectors are propagated in the graph, the unmasked label is also propagated in the graph. The masked labels not used in propagation are applied to compute loss and update model parameters. Masked Label Propagation works as label propagation but avoids label leakage and enhances the model's ability to generalize~\cite{shi2020masked}. Masked Label Propagation is implemented in popular GNN frameworks like PyG~\cite{fey2019pyg} and DGL~\cite{zheng2020distdgl}, and it was demonstrated to boost model accuracy on some datasets (e.g. OGB dataset~\cite{ogb-leaderboard}).
}

\begin{figure*}[h!]
  \begin{center}
    \includegraphics[clip,width=\textwidth]{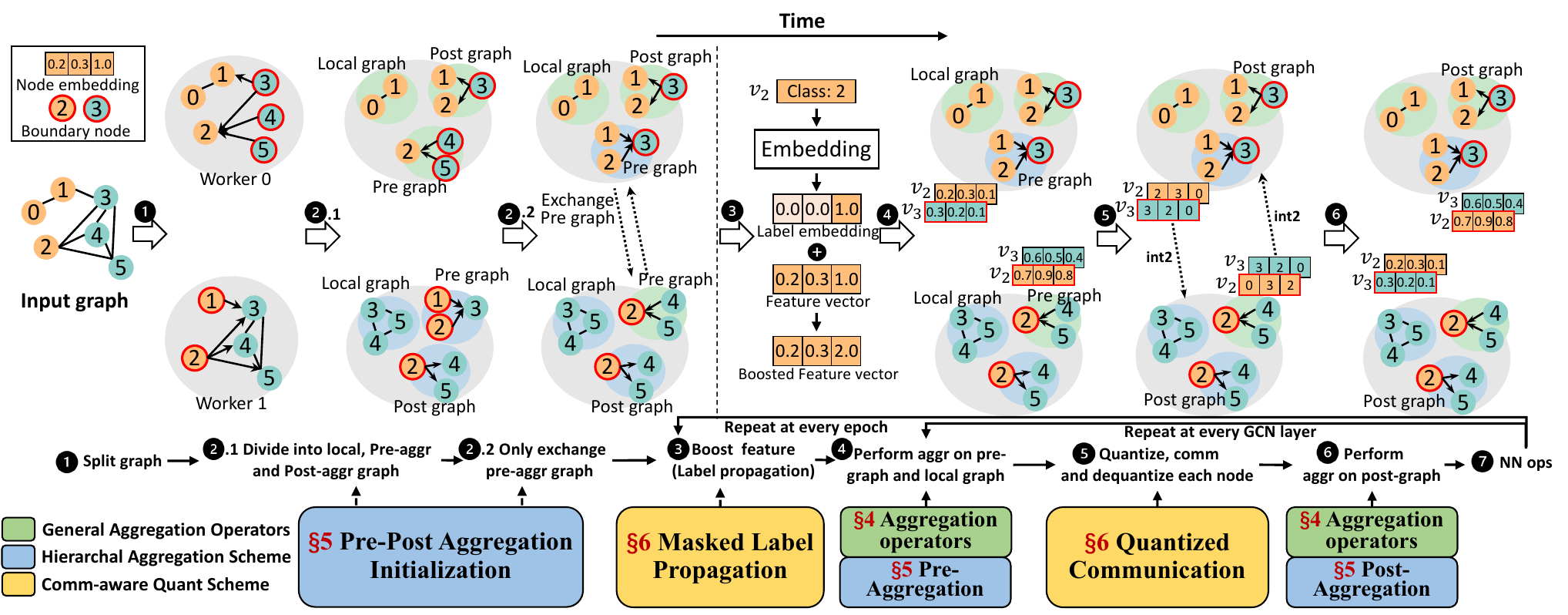}
    \caption{The flow of our proposed full-batch GCNs training system. Nodes with a red circle are boundary nodes obtained from other workers.}
    \label{fig/method:overview_procedure}
  \end{center}
\end{figure*}

{

\color{black}
\section{SuperGCN}\label{sec:superGCN}
\subsection{Design Methodology}
\method{} aims to minimize training time while maintaining high accuracy. \method{} consists of three components:
\setlength{\leftmargini}{10 pt}
\begin{itemize} 
\item\textbf{General and Efficient Aggregation Operators (Section~\ref{sec:node_level_opt}):} Optimizes node-level performance. 

\item\textbf{Hierarchical Aggregation Scheme (Section~\ref{sec:fram_work}):} Optimizes scalability, by reducing the communication volume, without affecting the algorithm or accuracy. 

\item\textbf{Communication-Aware Quantization Scheme (Section~\ref{sec:codesign}):} Employs aggressive quantization of communication to further improve scalability while preserving accuracy with algorithm design, as validated theoretically (Sec.~\ref{sec:quan_acc_ana}) and empirically (Sec.~\ref{sec:quan_acc_exp}).
\end{itemize}

\subsection{System Overview}
Fig.~\ref{fig/method:overview_procedure} shows \method{}'s components and their interactions:

\circled{1} Partition input graph into subgraphs using METIS~\cite{karypis1997parmetis}.
\circled{2} Split each subgraph into local, remote (pre-/post-aggregation) graphs based on communication reduction strategy (Sec.~\ref{subsec:pre-post-aggregation}). Exchange pre-aggregation graph between workers.
\circled{3} Embed randomly selected nodes' labels to their features (Sec.~\ref{sec:quantized_communication}).
\circled{4} Aggregate boundary nodes' features with pre-aggregation graph and local inner nodes' features with the local graph using optimized operators (Sec.~\ref{sec:node_level_opt}).
\circled{5} Quantize and transmit boundary nodes features (overlapped with local aggregation). Dequantize received features (Sec.~\ref{sec:quantized_communication}).
\circled{6} Aggregate boundary nodes' features with post-aggregation graph (Sec.~\ref{sec:node_level_opt}).
\circled{7} Apply NN operations (e.g., linear transform). Repeat from step \circled{4} for the next layer or \circled{3} for the next epoch.


\zhuanghpdc{For message-passing-based GNN models such as GAT~\cite{velickovic2017graph}, GIN~\cite{xu2018powerful}, and GraphSAGE~\cite{hamilton2017inductive}, their main difference lies in how they calculate the weights of neighbors (for example, GAT computes the weights through a self-attention mechanism). However, the core operation of these models remains neighbor aggregation (consisting of local aggregation and remote aggregation (communication)). \method{} primarily optimizes neighbor aggregation. Consequently, our system can be seamlessly applied to the distributed training of these message-passing-based GNN models.}

}
{
{\color{black}
\section{General and Efficient Aggregation Operators }\label{sec:node_level_opt}
At the node-level, full-batch GCN training uses two aggregation operators: \emph{Index\_add} and \emph{SpMM}, which are similar to each other, to aggregate the neighbors' feature vectors. The following section uses \emph{Index\_add} as an example.

In \emph{Index\_add}, rows in $src$ are added to $dst$ at positions specified by $idx$ (Fig.~\ref{fig/method:index_add_origin}). Unordered $idx$, incurred by the random connectivity of the graph, leads to a poor memory access pattern. We improve \emph{Index\_add}'s memory access pattern and locality:
\setlength{\leftmargini}{10 pt}
\begin{enumerate}
\item \emph{Clustering and sorting:} Sort $idx$ and cluster $src$ rows that aggregate to the same $dst$ row (Fig.~\ref{fig/method:index_add_sorted}). Though this might disrupt the access order of $src$ rows, since each $src$ row is still accessed sequentially and only once, the performance impact is minimal compared to improving $dst$ reuse.
\item \emph{Loop reordering:} Adjust memory access order on $src$ to enable register-optimized kernel for data reuse on $dst$.
\item \emph{Vector register-optimized kernel:} A shape-adaptive (aligned with cache line size) inner kernel with template-based code generation (Fig.~\ref{fig/method:index_add_reg}) to enhance memory access throughput and improve data reuse. 
\end{enumerate}
We also employ other optimizations including 2D dynamic parallelism and FLOPS-based load balancing (Fig.~\ref {fig/method:index_add_parallel}) to boost performance further. 
}

\begin{figure}[t]
    \centering
    \vspace{-1mm}
    \subfigure[Baseline vanilla implementation of index\_add.]{
        \begin{minipage}[b]{0.485\textwidth}
            \includegraphics[width=1\textwidth]{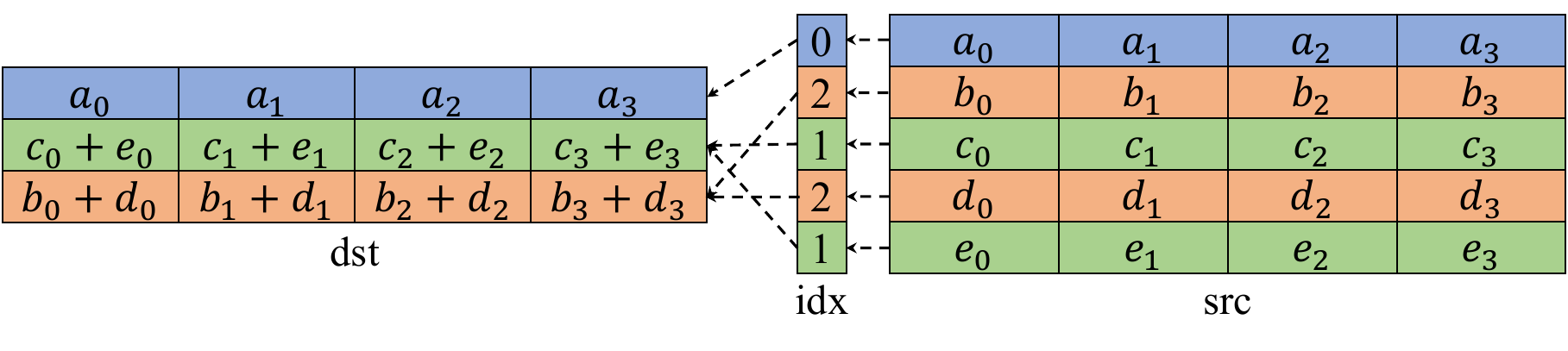}
        \end{minipage}
        \label{fig/method:index_add_origin}
    } 
    \vspace{-1mm}
    \subfigure[\zhuang{Cluster} $src$ based on \zhuang{sorted} $idx$ to improve memory access of $dst$.
    ]{
        \begin{minipage}[b]{0.485\textwidth}
        \includegraphics[width=1\textwidth]{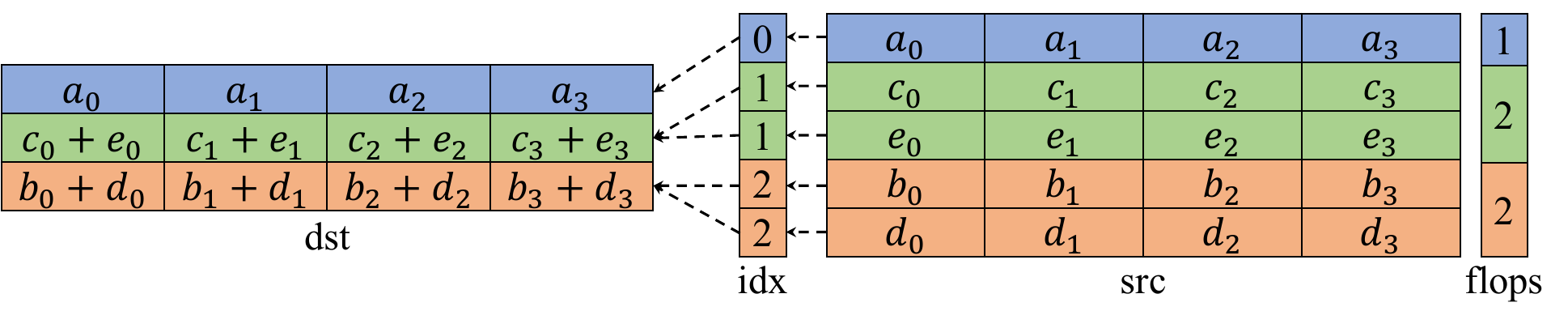}
        \end{minipage}
    \label{fig/method:index_add_sorted}
    } 
    \vspace{-1mm}
    
    \subfigure[Invoking a kernel optimized for vector register reuse \zhuang{on $dst$}. 
    ]
    {
		\begin{minipage}[b]{0.485\textwidth}
			\includegraphics[width=1\textwidth]{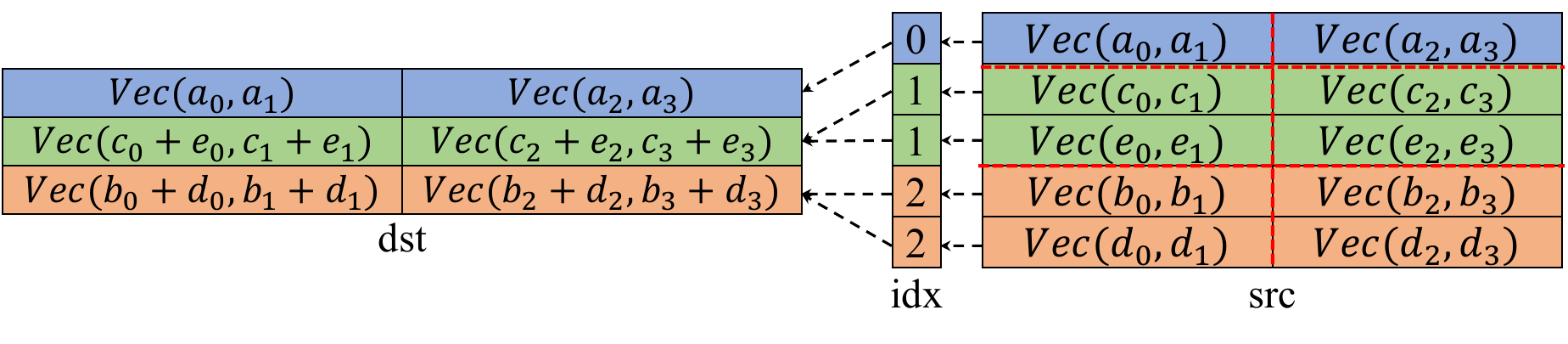}
		\end{minipage}
		\label{fig/method:index_add_reg}
	} \vspace{-1mm}

    \label{fig:hor_2figs_1cap_2subcap}
    \vspace{-1mm}
    \subfigure[
    A 2D dynamic parallelism strategy to determine the thread number on each dimension and improve load balance.
    ]{
        \begin{minipage}[b]{0.485\textwidth}
        \includegraphics[width=1\textwidth]{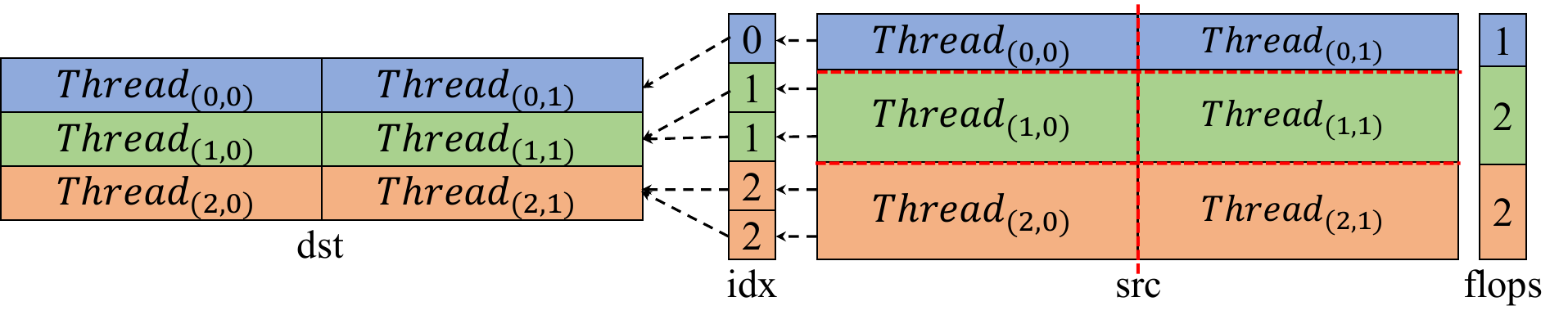}
        \end{minipage}
    \label{fig/method:index_add_parallel}
    } \vspace{-1mm}
    \caption{Steps of the proposed algorithm for optimizing the neighbor aggregation operator \emph{index\_add} on a single CPU.}
\end{figure}



\section{Hierarchical Aggregation Scheme}

\label{sec:proposed_methods}\label{sec:fram_work}

In distributed full-batch GCN training, communication performance is mainly affected by graph partitioning and the aggregation method.
In this section, we first describe our selected graph partitioning algorithm (Sec.~\ref{sec:graph_part}),  followed by our novel \emph{pre- and post-aggregation} method that reduces communication overhead based on partitioning results (Sec.~\ref{sec:dist_level_opt}). We then analyze the relationship between \emph{pre- and post-aggregation} and minimum vertex cover problem (Sec.~\ref{sec:minimum_vertex_cover}), and provide a performance model characterizing communication costs (Sec.~\ref{sec:permodel_basic}).


\subsection{Graph partitioning algorithm selection}
\label{sec:graph_part}
\zhuanghpdc{Regarding graph partitioning, we employ METIS~\cite{karypis1997parmetis} min-cut algorithm, which is widely adopted in popular GNN frameworks such as DGL~\cite{zheng2020distdgl}, PyG~\cite{fey2019pyg}, and so on~\cite{wan2022bns, wan2022pipegcn, peng2022sancus, ma2019neugraph, cai2021dgcl, wan2023adaptive, zhang2024sylvie}. METIS minimizes the total communication volume by minimizing the total number of cut edges. Additionally, 
it maximizes communication locality (neighboring subgraphs have higher communication volume). This locality property is beneficial for computing clusters and supercomputers where intra-node communication is significantly faster than inter-node communication.}

\subsection{\zhuang{Pre- and Post-aggregation}}
\label{subsec:pre-post-aggregation}
\label{sec:dist_level_opt}

\subsubsection{Constructing Pre- and Post-aggregation Subgraph}
\begin{figure}[t]
    \centering
    \subfigure[$S_0$'s remote graph. 
    The communication volume (dotted line) is 5.
    ]{
        \begin{minipage}[b]{0.2\textwidth}
            \includegraphics[width=1\textwidth]{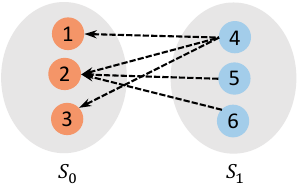}
        \end{minipage}
        \label{fig/method:complicated_original_graph}
    } \hspace{2mm}
    \subfigure[Convert to Pre-aggr graph. 
    The communication volume (dotted line) is 3.~\cite{md2021distgnn}
    ]{
        \begin{minipage}[b]{0.2\textwidth}
        \includegraphics[width=1\textwidth]{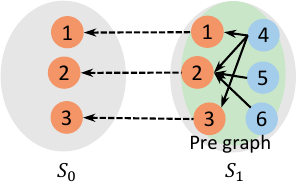}
        \end{minipage}
    \label{fig/method:complicated_pre_graph}
    } 
    \\
    \subfigure[Convert to Post-aggr graph. 
    The communication volume (dotted line) is 3.~\cite{wan2022bns, wan2023adaptive, mostafa2022sequential, wan2022pipegcn}
    ]{
        \begin{minipage}[b]{0.2\textwidth}
            \includegraphics[width=1\textwidth]{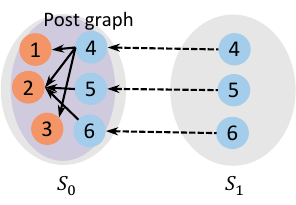}
        \end{minipage}
        \label{fig/method:complicated_post_graph}
    } \hspace{2mm}
    \subfigure[Ours : \zhuang{hybrid} of post-aggr and pre-aggr graph. 
    The communication volume (dotted line) is 2.
    ]{
        \begin{minipage}[b]{0.2\textwidth}
        \includegraphics[width=1\textwidth]{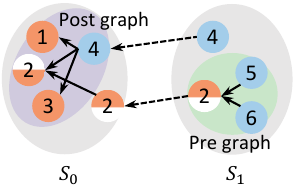}
        \end{minipage}
    \label{fig/method:complicated_optimal_graph}
    } 
    \caption{Strategies to construct a remote graph.}
    \label{fig/method:complicated_graph}
\end{figure}
After graph partitioning, each worker is allocated a subgraph. Each subgraph consists of two components: a local graph for message passing on inner nodes, and a remote graph for message passing on boundary nodes received from other subgraphs through the cut edges. To reduce communication overhead, the remote graph can be transformed into either a post-aggregation or pre-aggregation graph.
\emph{Pre} denotes performing the aggregation operation before communication, while \emph{post} refers to performing the aggregation operation after communication. For instance, consider $S_0$'s perspective of the remote graph shown in Fig.~\ref{fig/method:complicated_graph}. 
In this scenario, the communication volume along the cut edges (dotted line) is 5 nodes' features. By converting the remote graph to either the pre-aggr graph (Fig.~\ref{fig/method:complicated_pre_graph}) or the post-aggr graph (Fig.~\ref{fig/method:complicated_post_graph}, \zhuangppopp{the communication volume is reduced from 5 node features to 3 node features.}
\zhuang{To the best of our knowledge, most existing solutions choose either the pre-aggregation method~\cite{md2021distgnn} or the post-aggregation method \cite{wan2022bns, wan2023adaptive, mostafa2022sequential, wan2022pipegcn}.}

However, the two above solutions are suboptimal.
To achieve a more aggressive reduction in communication volume, we propose a simple yet effective method that categorizes each edge in the remote graph into either the pre-aggr graph or the post-aggr graph in Algo.~\ref{alg:pre-post-aggregation}. \zhuang{Initially, we treat the remote graph as a bipartite graph and identify all connected components in this graph. Subsequently, we introduce the minimum vertex cover algorithm to find the nodes that cover all cut edges (line 1-3 in Algo.~\ref{alg:pre-post-aggregation}). In Fig. \ref{fig/method:complicated_original_graph}, nodes $2$ and $4$ form the minimum vertex cover set. Then we traverse all edges in the remote graph (cut edges) (line 4-8 in Algo.~\ref{alg:pre-post-aggregation}). If the source node of the cut edges belongs to the minimum vertex cover (node $4$ and its cut edges), it's better to delay aggregation on these edges until the source node reaches the destination worker, as performing aggregation first would involve the transfer of multiple destination nodes and cause redundant communication. Conversely, when the destination node of cut edges is part of the minimum vertex cover (node $2$ and its cut edges), aggregation on these edges occurs prior to the transfer of results to the destination worker can reduce redundant communication (pre-aggr). Based on this strategy, a remote graph is transformed into a hybrid of pre-aggr graph and post-aggr graph, as shown in Fig.~\ref{fig/method:complicated_optimal_graph}.}

\begin{algorithm}[t]
      \small
\caption{\zhuang{A method for transforming a remote graph into a hybrid of pre- post-aggregation graph.}}
\label{alg:pre-post-aggregation}   

\KwIn{list of edges in remote graph $r\_edges$}
\KwOut{list of edges in pre-aggr graph $pre\_edges$, 
        list of edges in post-aggr graph $post\_edges$} 
        
    $r\_edges$ {$\gets$} to\_bipartite($r\_edges$)\ \ \ \ $v\_cover$ {$\gets$} set()

    \For{$subgraph$ in find\_connected\_components($r\_edges$)} {

        $v\_cover$.add(minimum\_vertex\_cover($subgraph$))
    
    }

    \For {$edge$ in $r\_edges$} {

        \If {$edge.src$ in $v\_cover$} {

            $post\_edges$.add($edge$)
        
        }

        \Else { 

            $pre\_edges$.add($edge$) 
        
        }
    
    }

    
        
        
        

\end{algorithm}

\subsubsection{Performing Pre- and Post-aggregation}
Our proposed approach involves the aggregation and communication stages of boundary nodes, including:
\setlength{\leftmargini}{15 pt}
\begin{enumerate}
    \item Perform pre-aggregation on node $5,6$ at $S_1$ to obtain the partial result of node $2$. 
    \item Send node $4$ and the partial result of node $2$ to $S_0$.
    \item Conduct post-aggregation on received node $4$ and the partial result of node $2$ at $S_0$ to get the aggregation result of node $1,2,3$.
\end{enumerate}
With our proposed strategy, the communication volume can be shrunk from 3 nodes to 2 nodes.

\subsection{Constructing Pre- and Post-aggregation Subgraph is Finding Minimum Vertex Cover}
\label{sec:minimum_vertex_cover}
\zhuanghpdc{To minimize communication overhead, we formulate the problem as a Minimum Vertex Cover (MVC) in a bipartite graph. This section explains how MVC ensures an optimal communication volume.}

\subsubsection{Minimum Vertex Cover (MVC) in bipartite graphs}
In a bipartite graph $G = (U, V, E)$, a vertex cover is a subset of vertices $C \subseteq U \cup V$ such that every edge in $E$ has at least one endpoint in $C$. The \textbf{minimum vertex cover} is the smallest such set. König's theorem~\cite{konig1931graphen} establishes that in bipartite graphs, the size of a minimum vertex cover equals the size of a maximum matching, allowing polynomial-time solutions.

\subsubsection{Achieving Optimal Communication Volume via Minimum Vertex Cover}
\begin{figure}[t!]
    \centering
    \includegraphics[clip,width=0.48\textwidth]{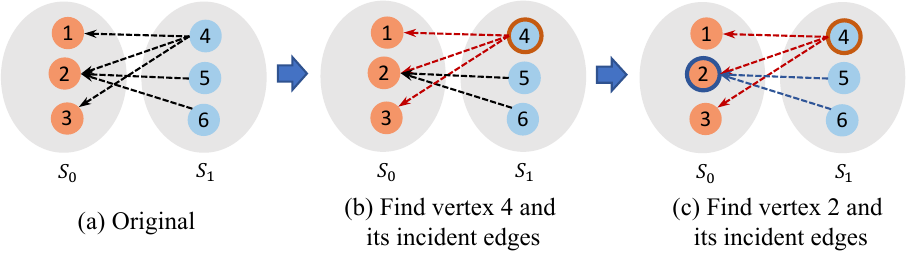}
    \caption{Achieving optimal communication volume via finding minimum vertex cover.}
    \label{fig/method:minimum_vertex_cover}
\end{figure}
Let the remote graph (Fig.~\ref{fig/method:complicated_original_graph}) induce a bipartite graph \(G = (U, V, E)\), where each edge \(e = (u, v) \in E\) represents a data transfer between vertices \(u\) and \(v\). If data were transferred along every edge, a vertex connected to multiple edges might be transmitted repeatedly, leading to redundancy. To eliminate this redundancy, we seek a set of vertices \(C \subseteq U \cup V\) such that every edge is “covered” by at least one vertex in \(C\); mathematically, we require that
\[\footnotesize
\forall (u, v) \in E, \quad u \in C \quad \text{or} \quad v \in C.
\]

This set \(C\) effectively “absorbs” the data transfers for all its incident cut edges, meaning that transferring the data corresponding to the vertices in \(C\) is sufficient to account for all communications. Our goal is to minimize the communication volume, which is directly proportional to the number of vertices transmitted. Thus, we formulate the problem as
\begin{equation}\footnotesize
\begin{aligned} \label{eq:minimum_vertex_cover}
\min_{C \subseteq U \cup V} |C| \quad \text{subject to} \quad \forall (u,v) \in E, \; u \in C \text{ or } v \in C.
\end{aligned}
\end{equation}

Take Fig.~\ref{fig/method:minimum_vertex_cover}(a) as an example, the goal is to find a vertex set (vertex 2 and 4) that covers all edges while minimizing the number of vertices. Formulation~\ref{eq:minimum_vertex_cover} is exactly the definition of the MVC problem~\cite{diestel2017graph} on a bipartite graph. So we conclude that \textbf{\textit{finding the optimal communication volume is equivalent to solving the MVC problem on \(G\)}}. The optimal communication volume corresponds to \(|\text{MVC}|\): the minimal number of vertices needed to cover all edges. Also, the MVC problem on bipartite graphs has an optimal solution (by König’s theorem~\cite{konig1931graphen} and Hopcroft–Karp algorithm~\cite{hopcroft1973n}). 

After identifying the minimum vertex cover set \(C\) (e.g., vertex 2 and 4 in Fig.~\ref{fig/method:minimum_vertex_cover}(c)), we construct pre- and post-aggregation graph based on \(C\), transforming the original edges communication into vertices communication on \(C\), thereby achieving the optimal communication volume.

\subsection{Communication Performance Model}
\label{sec:permodel_basic}
\wu{After partitioning the graph and aggregating the nodes, the communication cost required in distributed full-batch GCN training can be estimated. We use ${Comm^{i, j}}$ to represent the communication volume from process $i$ to process $j$. For distributed training with $P$ processes, the communication time $T_{comm}$ is expressed as follows:} 
\begin{equation}\footnotesize
\label{eq:comm_volume}
\begin{aligned}
T_{comm}^{i, j} &= \frac{Comm^{i, j} \times BIT_{fp32}}{BW_{comm}} + L_{comm} \\
T_{comm} &= \max_{i=0}^{P}(\sum_{j=0}^{P}{T_{comm}^{i, j}}) 
\end{aligned}
\end{equation}


\zhuanghpdc{
The upper equation computes the communication time as the sum of the data transfer time, given by \(\frac{V_{comm}^{i, j}}{BW_{comm}}\), and the fixed communication latency \( L_{comm} \).  
Since the communication pattern in distributed full-batch GCN training is highly unbalanced, to account for this imbalance, the lower equation selects the communication time of the process with the longest communication time (serves as the bottleneck) as the global communication time \( T_{comm} \).
}

\section{Communication-Aware Quantization Scheme}
\label{sec:codesign}





\label{sec:quantized_communication}
\lingqi{To further reduce the communication volume,} in this section we introduce a communication-aware quantization scheme that leverages quantization in the communication of distributed full-batch GCN training, while maintaining model accuracy. \lingqi{We illustrate the workflow of our scheme in Sec.~\ref{sec:workflow}. We then provide a thorough performance analysis in Sec.~\ref{sec:perf_analysis} and accuracy analysis in Sec.~\ref{sec:quan_acc_ana}.}


\subsection{Workflow of quantization communication}
\label{sec:workflow}
The workflow of our scheme (as shown in Fig.~\ref{fig/method:quant_timeline}) is 

\begin{figure}[t!]
    \centering
    \includegraphics[clip,width=0.48\textwidth]{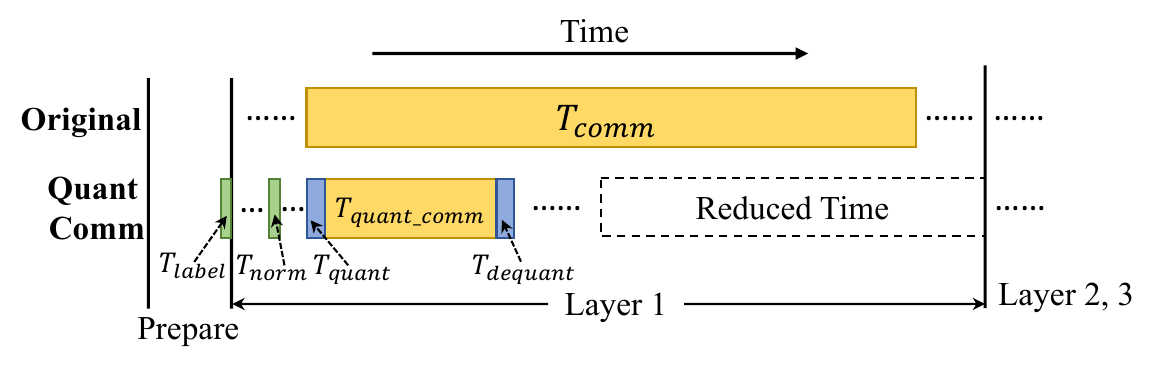}
    \caption{Traditional communication (top) and our communication quantization scheme (bottom) in GCN training.}
    \label{fig/method:quant_timeline}
\end{figure}

\setlength{\leftmargini}{15 pt}
\begin{enumerate}
    \item \textbf{Masked label propagation.} At the beginning of each epoch, we convert the labels of randomly selected nodes into embedding and add them to the initial feature vectors, which allows the labels to propagate during neighbor aggregation (\circled{3} in Fig.~\ref{fig/method:overview_procedure}).
    Unselected labels are used to update model parameters.
    \item \textbf{Normalization.} Large-magnitude outliers introduce large quantization errors. Therefore, we apply LayerNormalization on the embedding table before each GCN layer to remove outliers and smoothen the data distribution. 
    \item \textbf{Quantize, comm and dequantize.}
    \zhuanghpdc{Quantize the embedding table with quantization parameters (zero-point and scale). Transfer the quantized embedding table and quantization parameters to other workers. After communication, dequantize received embedding tables with the received quantization parameters. (\circled{5} in Fig.~\ref{fig/method:overview_procedure})}
    
    
    
\end{enumerate} 

\subsection{Performance analysis}
\label{sec:perf_analysis}

\jiajun{In this section, we present the performance improvements achieved by our communication-aware quantization scheme through theoretical analysis. The key takeaway is that our approach effectively reduces communication volume and enhances communication efficiency.}


\subsubsection{Performance model of our scheme.} 

\zhuanghpdc{Quantization reduces the communication volume yet requires additional overhead to preserve model accuracy. Accounting for these overheads, we restructure our performance model (Eqn~\ref{eq:comm_volume}) into three components:}

1) $\boldsymbol{T_{pre\_quant}}$: Time for Masked label propagation at the start of each epoch and LayerNormalization before each GCN layer. These operations are performed on the local subgraph and do not increase the communication volume:
\begin{equation}\footnotesize
\label{eq:T_prequant}
\begin{aligned}
T_{pre\_quant}^{i} &= \frac{SubGraph^{i} \times BIT_{fp32}}{TH_{cal}}
\end{aligned}
\end{equation}

2) $\boldsymbol{T_{quant}}$ and $\boldsymbol{T_{dequant}}$: Time for quantizing and dequantizing the communication data. We support quantizing the FP32 data into intX (where X can be 2, 4, or 8 bits). The communication volume can be reduced by a factor of $\frac{BIT_{fp32}}{BIT_{intX}}$, but this reduction introduces additional computational cost:
\begin{equation}\footnotesize
\label{eq:T_quant_and_dequant }
\begin{aligned}
T_{quant}^{i,j} = T_{dequant}^{j,i} &= \frac{Comm^{i,j} \times (BIT_{fp32} + BIT_{intX})}{TH_{cal}}
\end{aligned}
\end{equation}

3) $\boldsymbol{T_{quant\_comm}} \textbf{ with Params}$: Quantization introduces quantization-related parameters (zero-point and scale), which are stored as FP32 values. These parameters need to be transferred to other workers for dequantization. Therefore, the time for quantized communication (transferring data and parameters) from process $i$ to process $j$ is updated as:
\begin{equation}\footnotesize
\label{eq:T_para}
\begin{aligned}
T_{quant\_comm}^{i,j} &= \frac{(Comm^{i, j} \times BIT_{intX}) + (\boldsymbol{Params}^{i, j} \times BIT_{fp32})}{BW_{comm}} + L_{comm}
\end{aligned}
\end{equation}

Taking all these components into account, the total quantized communication time is estimated to be:
\begin{equation}\footnotesize
\label{eq:T_quant_total}
\begin{aligned}
T_{quant\_comm} &= \max_{i=0}^{P}(T_{pre\_quant}^{i} + \sum_{j=0}^{P}{(T_{quant}^{i,j} + T_{quant\_comm}^{i, j} + T_{dequant}^{i,j})}) 
\end{aligned}
\end{equation}

\subsubsection{Performance gain of our scheme.} \label{sec:perf_gain}In this section, we analyze the performance gain of our scheme under different scales (bottlenecks in throughput or latency). To facilitate analysis, we define four ratios $\alpha$, $\beta$, $\gamma$ and $\delta$ where 
\begin{equation}\footnotesize
\label{eq:T_ab}
\begin{aligned}
\alpha &= \frac{Comm^{i,j}}{Params^{i,j}} \sim O(10^2) \\ 
\beta &= \frac{TH_{cal}}{BW_{comm}} \sim O(10^2) \\
\gamma &= \frac{BIT_{fp32}}{BIT_{intX}} = \frac{32}{X} \\ 
\delta &= L_{comm} : \frac{Comm^{i,j} \times BIT_{intX}}{BW_{comm}}
\end{aligned}
\end{equation}
where \(\alpha\) represents the ratio of communication data volume to quantization-related parameters volume, \(\beta\) is the ratio of computing throughput to communication bandwidth of the hardware, \(\gamma\) is the ratio of bit-width before and after quantization, and \(\delta\) is the ratio of communication latency to data transfer time. The performance gain of our scheme can be expressed as follows:
\begin{equation}\footnotesize
\label{eq:speedup1}
\begin{aligned}
Speedup &= \frac{T_{comm}}{T_{quant\_comm}} = \frac{\alpha\beta(\gamma+\delta)}{(1+\delta)\alpha\beta+2\alpha(1+\gamma)+\beta\gamma} \approx \frac{\gamma+\delta}{1+\delta}
\end{aligned}
\end{equation}
Note that \( T_{\text{pre\_quant}} \) is omitted in Equation 8, as it is negligible compared to the communication time with processes \( P \).
\textbf{In the case of medium scales, communication is throughput-bound.} This implies that $\delta \to 0$. 
Our scheme reduces communication volume $\gamma$ times and can achieve nearly $\gamma$ speedup (left side in Fig.~\ref{fig/method:perf_analysis}). For example, with Int2 quantized communication, the speedup approaches 16$\times$.
\textbf{As the number of processes ($P$) increases, the communication bottleneck gradually changes to latency.} This is reflected in the increase of $\delta$ until $\delta$ approaches infinity ($\delta \to \infty$). According to Eqn~\ref{eq:speedup1}, the performance improvement brought about by our scheme diminishes as the system increases the scale, yet it does not have any negative impact. Furthermore, our scheme can make the communication shift to latency bound with fewer processes ($P'$), resulting in an absolute time reduction of $(P-P')\cdot L_{comm}$ compared to the case without our scheme (as shown in Fig.~\ref{fig/method:perf_analysis}).

\begin{figure}[t!]
    \centering
    \includegraphics[clip,width=0.38\textwidth]{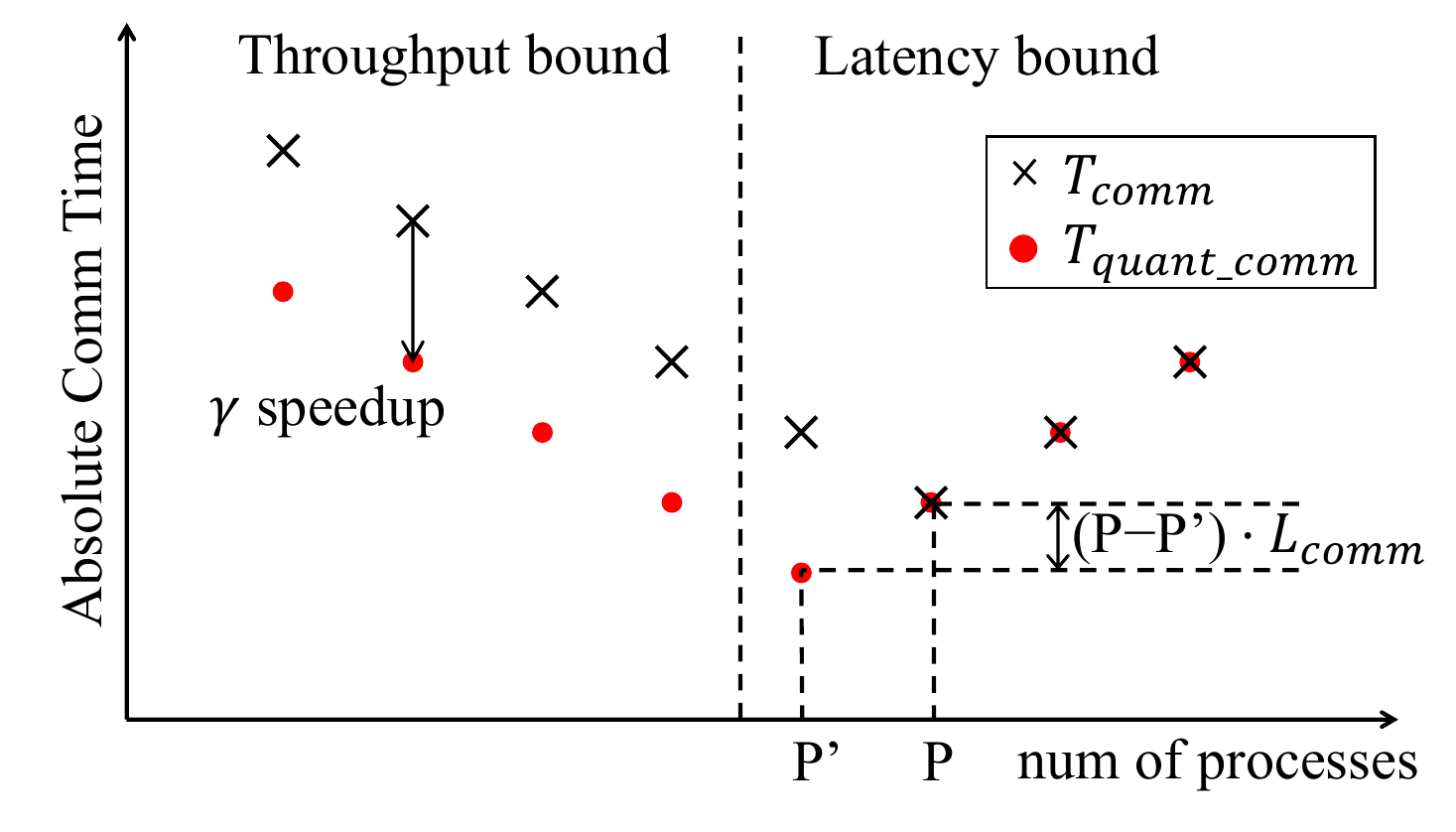}
    \caption{Performance analysis of quantized communication.}
    \label{fig/method:perf_analysis}
\end{figure}

\subsection{Accuracy analysis}
\label{sec:quan_acc_ana} 
In this subsection, we analyze the methods we use from the perspective of model accuracy.

\textbf{Lemma 1.} With the GNNs model, given a global minimum of loss $\mathcal L(W^*)$, we make the following assumptions:

(1) the gradient $\nabla \widetilde {\mathcal L}(W_t)$ is $\rho$-Lipschitz continuous, (2) the approximate gradient is unbiased, {\small $\mathbb{E}[\nabla \widetilde {\mathcal L}(W_t)] = \nabla {\mathcal L}(W_t)$}, (3) the variance of the approximate gradient is bounded, 
{\small$Var[\lVert \nabla \widetilde {\mathcal L} (W_t) - \nabla \mathcal L(W_t) \rVert_2] = \mathbb{E}[\lVert \delta_t \rVert_2] \leq K$}, $K$ is a constant and $K > 0$.

Then select any t in epoch{\small \{1,...,T\}}, we have 
\begin{equation}\footnotesize
\begin{aligned}
\mathbb{E}[\lVert \nabla \widetilde {\mathcal L} (W_t) \rVert_2^2] \leq \frac{2 ({\mathcal L} (W_1) - {\mathcal L} (W^*))}{T(2\eta-\rho\eta^2)} + \frac{\eta\rho}{2-\eta\rho} \cdot K^2 
\end{aligned}
\end{equation}

The proof of Lemma 1 follows the analysis in~\cite{wan2023adaptive}, adapted to our setting. The first term of the right side will go to $0$ while {\small$T \rightarrow \infty$}. This means under the quantization scheme we maintain {\small$O (\frac{1}{T})$} convergence rate, and model will converge to the neighborhood of a stationary point, the radius of the neighborhood is determined by the gradient variance $K$.



\textbf{Lemma 2.} With the GCNs model, by embedding label into the vector space of feature, label propagation and feature propagation are approximately unified into message passing operations in GCNs. That is to say, we have:
\begin{equation}\footnotesize
\begin{split}
    H^{(l+1)} = AH^{(l)}W^{(l)} = A^{l}(X+Y_{embed})(W^{(0)}W^{(1)} \cdots W^{(l)}) \\
    = A^{l}X(W^{(0)}W^{(1)} \cdots W^{(l)}) +  A^{l}Y_{embed}(W^{(0)}W^{(1)} \cdots W^{(l)})
\end{split}
\end{equation}

where $H^{(l)}$ is the embedding after GCN layer $l$, $A$ is the adjacent matrix, $W^{l}$ is the weight matrix in GCN layer $l$, $X$ is the initial feature vectors, $Y_{embed}$ is the embedding from label, while $Y_{embed} = YW_{embed}$. 

\textbf{Proposition 1.} \zhuanghpdc{Lemma 1 shows that with quantized communication, the model still maintains the same convergence rate. According to Lemma 2, adding embedded labels to features can be treated as a type of label propagation. Based on \cite{shi2020masked}, integrating label propagation into feature propagation can enhance the connectivity between nodes with the same label. This enhanced connectivity brings the embeddings of nodes with the same label closer in the high-dimensional latent space, partially mitigating the issue where quantization shortens the distances between the embedding of nodes with different labels. This enables the model to more precisely separate nodes with different labels. Therefore, this enhances the model's tolerance to precision loss.}

\section{Implementation}
Our framework is built on top of PyTorch and PyTorch Geometric (PyG)~\cite{fey2019pyg}. Currently, PyG lacks support for distributed full-batch GNNs training. Consequently, we implement the entire training process including preprocessing, graph partitioning, subgraph construction, and so on. For communication primitives, we employ MPI\_Alltoallv. 

 \begin{table*}
      \caption{The graph datasets and model hyperparameters used in experiments.}        
      \centering
      \vspace{-4pt}
      \resizebox{1\linewidth}{!}{
          \begin{tabular}{ |l|c|c|c|c|c|c|c|c|c|c|c| }
          \hline
          \bf{Datasets} & \bf{\#Vertex} & \bf{\#Edges} & \bf{\#Feat} & \bf{\#Class} & \bf{\#Hidden} & \bf{\#Epochs} & \bf{Dropout} & \bf{Learning Rate} & \bf{Norm Type} \\
          \hline
          \textsf{Ogbn-arxiv} & 169,343 & 1,166,243 & 128 & 40 & 256 & 250 & 0.5 & 0.01 & LayerNorm \\
          \textsf{Reddit} & 232,965 & 114,615,892 & 602 & 41 & 256 & 250 & 0.5 & 0.01 & LayerNorm \\
          \textsf{Ogbn-products} & 2,449,029 & 61,859,140 & 100 & 47 & 256 & 250 & 0.5 & 0.01 & LayerNorm \\
          \textsf{Proteins} & 8,745,542 & 1,309,240,502 & 128 & 256 & 256 & 200 & 0.5 & 0.01 & LayerNorm \\
          \textsf{Ogbn-papers100M$^{1}$} & 111,059,956 & 1,615,685,872 & 128 & 172 & 256 & 200 & 0.5 & 0.005  & LayerNorm \\
          \zhuang{\textsf{Ogb-lsc-mag240M$^{2}$}} & 121,751,666 & 2,593,241,212 & 768 & 153 & 256 & 300 & 0.5 & 0.005 & LayerNorm \\
          \textsf{UK-2007-05} & 105,896,555 & 3,738,733,648 & 128 & 172 & 128 & 200 & 0.5 & 0.01 & LayerNorm \\
          \zhuangppopp{\textsf{IGB260M}} & 269,346,174 & 3,995,777,033 & 1024 & 19 & 256 & 200 & 0.5 & 0.01 & LayerNorm \\
          \hline
           \end{tabular}  
        }
         \vspace{-5pt}
\small
$^{1}$We convert this directed graph into an undirected graph on Fugaku. \quad $^{2}$We extract the undirected homogeneous papers citation graph
       \label{table/experiment:dataset_and_model} 
\end{table*}     

\begin{figure*}[ht]
    \centering
    \includegraphics[width=0.92\linewidth]{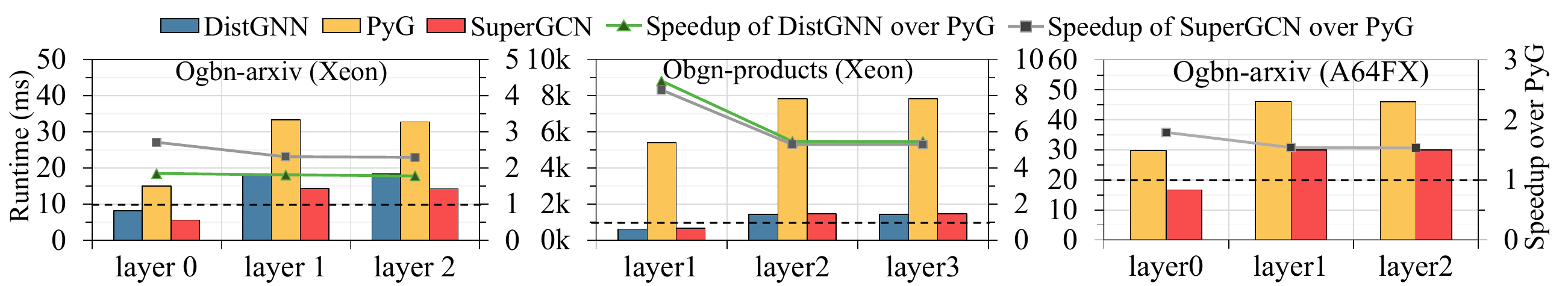}
    \caption{ Performance comparisons of aggregation operators for different datasets on a single Intel Xeon Gold 6148 processor and one CMG of an Arm Fujitsu A64FX processor. A64FX processor is divided into four CMGs, each appearing as a single NUMA node.}
    \label{fig/experiment:single_socket_combo}
\end{figure*}

\subsection{Single node level}
Unlike typical CPUs that are optimized for latency,  A64FX is a CPU optimized for throughput. It features a larger cache line size, higher memory bandwidth, more compute cores, wider SIMD instruction (Arm SVE 512), and longer operation latency. Therefore, our optimization goal is to provide more parallelism (e.g., aggressive software prefetch) to hide long latency. 
For x86 CPUs, we implement and optimize the aggregation operators with Intel AVX-512.

\subsection{Distributed level}
For graph partitioning, 
to ensure a balanced distribution of training samples and prevent load imbalances among different workers, we assign node weights based on in-degree of nodes and training masks when configuring METIS~\cite{karypis1997parmetis}.


For the minimum vertex cover in pre- and post-aggregation, we optimize the implementation of the NetworkX library~\cite{hagberg2008exploring}. This significantly reduces the preprocessing time.

\subsection{Quantization \lingqi{of Communication}}

In our implementation, we fix the bit-width of quantization to Int2. We implemented the quantization process based on~\cite{wan2023adaptive} and made further optimizations:

\zhuanghpdc{(1) Decentralized int2 quantization communication scheme. In our proposed scheme, each worker quantizes and dequantizes the message without any synchronization or communication with a centralized master.}

\zhuanghpdc{(2) Fusion of quantization parameters calculation and quantization kernels. To reduce redundant memory access, our fused kernel first retrieves 4 rows of the embedding table (necessary for packing four int2 values into one int8 for compatibility) and calculates quantization parameters for these rows. Then, it reuses data from cache and quantizes them using the newly calculated parameters.}

\zhuanghpdc{(3) Latency reduction in quantization kernel. We improve performance by replacing the long latency division operation (98 cycles on Fujitsu A64FX~\cite{A64FXManual}) with reciprocal estimate and multiplication operations and eliminating random number generation to shorten instruction dependency chains and improve throughput.}

\zhuanghpdc{(4) Vectorization for quantization and dequantization kernel on different CPUs. }

\section{Evaluation}
\label{sec:evaluation}

\subsection{\zhuang{Experimental Setup}}

\begin{figure*}[ht]
    \centering
    \includegraphics[width=1\linewidth]{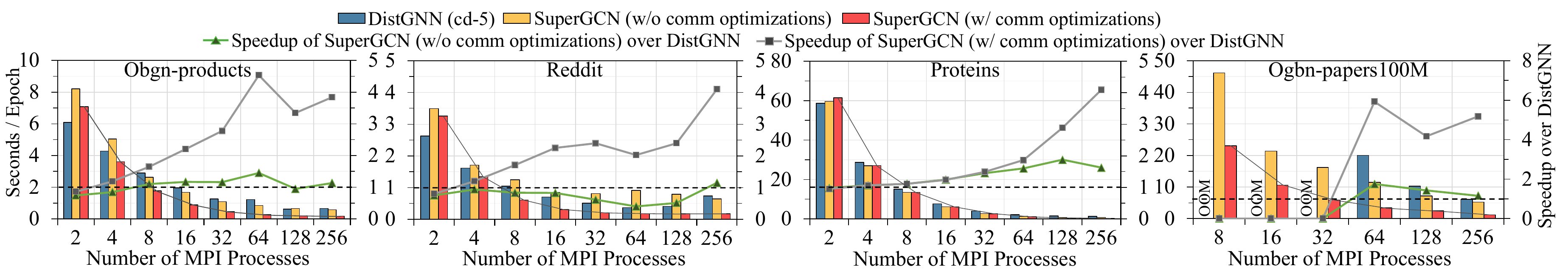}
    \caption{Performance comparison with DistGNN and scaling for different datasets on ABCI (Intel). OOM refers to the out-of-memory runs. The results in the bar chart correspond to the left y-axis (performance), while the results in the line chart correspond to the right y-axis (speedup).}
\label{fig/experiment:multi_socket_perf_combo_abci}
\end{figure*}
\begin{figure*}[ht]
    \centering
    \includegraphics[width=1\linewidth]{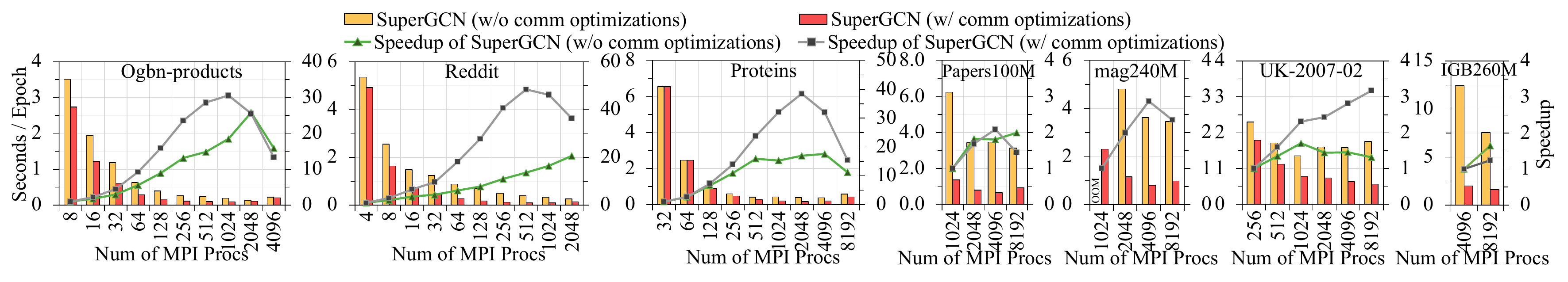}
    \caption{Performance and scaling for different datasets on Fugaku (Arm). The results in the bar chart correspond to the left y-axis (performance), while the results in the curves correspond to the right y-axis (speedup). The speedup is over the smallest number of MPI processes the dataset can be executed on. Each A64FX in Fugaku runs four MPI ranks: the number of processors used in each run becomes the number of MPI ranks / 4.}
    \label{fig/experiment:strong_scaling_fugaku}
\end{figure*}


\textbf{Hardware.} 
We conducted experiments on two supercomputers: ABCI~\cite{abci} and Fugaku~\cite{sato2020co}. ABCI's compute nodes are connected with InfiniBand and contain two Intel Xeon Gold 6148 CPUs, each with 20 cores.
Fugaku's compute nodes are connected via the Tofu interconnection network and feature Fujitsu Arm A64FX processors. These processors have 48 cores, organized into 4 CMGs with each CMG containing 12 cores. 
\textbf{Datasets.} Table~\ref{table/experiment:dataset_and_model} shows the graph datasets in the experiments. The datasets are a) part of the "Open Graph Benchmark" suite~\cite{hu2020open, hu2021ogb}; b) the widely used real dataset "Reddit"~\cite{hamilton2017inductive}; c) two large-scale graphs "Proteins"~\cite{azad2018hipmcl} and "UK-2007-05"~\cite{BoVWFI, BRSLLP};
d) the latest and largest publicly available dataset "IGB260M"~\cite{khatua2023igb}. 
\textbf{Models.} We use a three layers GraphSAGE~\cite{hamilton2017inductive} model 
for the experiments. GraphSAGE is the most commonly used GCN model to assess the performance of GNN frameworks.
The model settings for different datasets are shown in Table~\ref{table/experiment:dataset_and_model}. 
\textbf{Baselines.} We choose the state-of-the-art system \textit{DistGNN} as the baseline on ABCI. \textit{DistGNN} is optimized for Intel clusters. It proposes delayed communication to reduce communication overhead. Following their paper's suggestion, we set the delayed communication epoch number to 5 (cd-5). Since currently there is no proper baseline optimized for Arm processors, we report the performance of our framework on Fugaku as: (1) \textit{SuperGCN (w/o comm optimizations)}, which is our implementation before applying the proposed communication optimization, and (2) \textit{SuperGCN (w/ comm optimizations)}, which includes all the proposed communication optimization (pre-post aggregation and quantization communication). We also compare our methods with baselines optimized for full-batch distributed GCN training on GPU clusters. The results of these baselines are collected from their original papers. 

\subsection{Performance of Aggregation on a Single CPU}

%

This section reports the performance of the proposed aggregation operators as these operators are the most time-consuming of full-batch GCN training on a single CPU. Fig.~\ref{fig/experiment:single_socket_combo} shows the performance comparison of aggregation operators on a single Intel Xeon Gold 6148 processor. Due to the limitations of memory capacity, we only show results on a subset of the dataset. 
On the Xeon, \zhuang{\method{}} achieves a speedup of 1.8$\times$ to 8.4$\times$ on different model layers over PyG, and $\sim$1.22$\times$ over Intel's library DistGNN. 
When processing larger datasets, such as ogbn-products, \zhuang{\method{}} eliminates more redundant memory accesses. As a result, \zhuang{\method{}} achieves higher speedup on larger datasets. 
We also compare \zhuang{\method{}} with PyG on one CMG of an Arm Fujitsu A64FX processor ($\frac{1}{4}$ of the processor). We achieve a speedup ranging from 1.5$\times$ to 1.8$\times$ on the Ogbn-arxiv dataset. 

\begin{figure*}[ht] 
    \centering
    \includegraphics[width=1\linewidth]{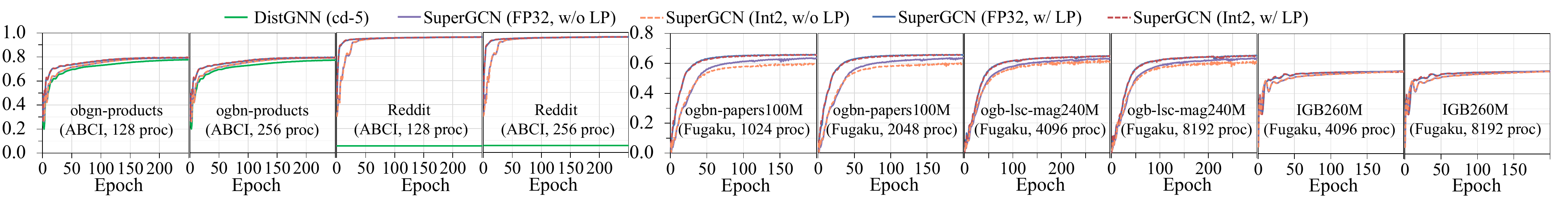}
    \caption{\zhuang{Accuracy of test datasets (products, Reddit, papers100M and IGB260M) and validation dataset (Ogb-lsc-mag240M). \zhuangppopp{W/o LP refers to training without masked label propagation, while w/ LP is to train with masked label propagation.}}}
    \label{fig/experiment:accuracy_combo_abci}
\end{figure*}

\begin{table*}[ht]
    \centering
        \caption{Accuracy of different datasets. The \emph{N/A} refers to the abnormal accuracy. \zhuang{SoTA accuracy of mini-batch (neighbor sampling) with the same model as SuperGCN is collected from Ogb-leaderboard~\cite{ogb-leaderboard} and~\cite{wan2022bns}~\cite{kaler2022accelerating}~\cite{kaler2023communication}~\cite{khatua2023igb}.}}
    \resizebox{\linewidth}{!}{      
    \begin{tabular}{|l|c|c|c|c|c|c|c|c|c|c|c|c|}
        \hline
        \multirow{2}*{\textbf{Method}} & \multicolumn{3}{c|}{\zhuang{\textbf{Obgn-products (ABCI)}}} & \multicolumn{3}{c|}{\zhuang{\textbf{Reddit (ABCI)}}} & \multicolumn{2}{c|}{\zhuang{\textbf{Ogbn-papers100M (Fugaku)}}} & \multicolumn{2}{c|}{\zhuang{\textbf{Ogb-lsc-mag240M (Fugaku)}}} & \multicolumn{2}{c|}{\zhuang{\textbf{IGB260M (Fugaku)}}} \\ \cline{2-13}
        & 64 procs & 128 procs & 256 procs & 64 procs & 128 procs & 256 procs & \zhuang{1024 procs} & \zhuang{2048 procs} & \zhuang{4096 procs} & \zhuang{8192 procs} & \zhuang{4096 procs} & \zhuang{8192 procs} \\ 
        \hline
        \textsf{DistGNN \zhuang{(cd-5)}} & 78.16 & 77.76 & 77.27 & N/A & N/A & N/A & - & - & - & - & - & -\\ 
        \textsf{\zhuangppopp{SuperGCN (FP32, w/o LP)}} & 79.14 & 79.15 & 79.17 & 96.14 & 96.12 & 96.16 & 63.58 & 63.62 & 63.33 & 63.33 & 54.90 & 54.89\\
        \textsf{\zhuangppopp{SuperGCN (Int2, w/o LP)}} & {79.42} & {79.24} & {79.33} & 96.10 & 96.17 & 96.16 & 60.19 & 60.42 & 62.73 & 60.91 & \textbf{54.96} & \textbf{54.96}\\
        \textsf{\zhuangppopp{SuperGCN (FP32, w/ LP)}} & 79.64 & 79.62 & \textbf{79.61} & \textbf{96.28} & \textbf{96.27} & 96.23 & 65.62 & \textbf{65.82} & \textbf{64.99} & \textbf{65.29} & 54.39 & 54.37 \\
        \textsf{\zhuangppopp{SuperGCN (Int2, w/ LP)}} & \textbf{79.68} & \textbf{79.62} & 79.26 & 96.26 & \textbf{96.27} & \textbf{96.27} & \textbf{65.71} & 65.70 & 64.87 & 65.23 & 54.51 & 54.93 \\
        \hline
        \textsf{\zhuang{Mini-batch (GraphSAGE)}} & \multicolumn{3}{c|}{78.70~\cite{ogb-leaderboard}} & \multicolumn{3}{c|}{95.40~\cite{wan2022bns}} & \multicolumn{2}{c|}{64.91~\cite{kaler2022accelerating}} & \multicolumn{2}{c|}{65.10~\cite{kaler2023communication}} & \multicolumn{2}{c|}{54.95~\cite{khatua2023igb}} \\ 
        \hline
    \end{tabular}
    }
    \label{table/experiment:accuracy}
\end{table*}

\subsection{Performance and Scaling on Multiple CPUs }

\textbf{ABCI supercomputer}. We conduct a performance and scaling comparison between \method{} and DistGNN on multiple CPUs using ABCI (Xeon CPUs) across various datasets. Fig.~\ref{fig/experiment:multi_socket_perf_combo_abci} illustrates the performance and speedup achieved over DistGNN on four different datasets. Due to the memory capacity limitation, we begin with 8 MPI processes for ogbn-papers100M. \method{} achieves speedups ranging from 0.9$\times$ to 6.0$\times$ compared to DistGNN. The results show that speedup becomes more significant as the number of processes increases, due to communication becoming the main bottleneck. These experimental results confirm the efficacy of our approach in reducing communication overhead. It is worth noting that those speedups are on the Xeon-based supercomputer, for which DistGNN is particularly designed. 

The scaling of \zhuang{\method{} (w/o and w/ comm optimizations)} and DistGNN on ABCI are shown in Fig.~\ref{fig/experiment:multi_socket_perf_combo_abci}. DistGNN suffers from out-of-memory issues on Obgn-papers100M. Across all datasets, \zhuang{\method{} (w/ comm optimizations)} demonstrates superior strong scalability when compared with DistGNN. Notably, on Ogbn-papers100M, \zhuang{\method{}} achieves a speedup that is close to linear scaling. On 
proteins, \zhuang{\method{}} even obtains a superlinear speedup. 

\textbf{Fugaku supercomputer.} Fig.~\ref{fig/experiment:strong_scaling_fugaku} shows the performance and scaling of \zhuang{\method{}} on Fugaku. These results demonstrate the scalability of \zhuang{\method{}} on Arm-based supercomputers. In summary, \method{} scales to a maximum of 8,192 MPI ranks, especially on the largest publicly available dataset IGB260M. \emph{To the best of our knowledge, this stands as the highest scalability achieved by a full-batch GNNs training system.}

Furthermore, as shown in Fig.~\ref{fig/experiment:strong_scaling_fugaku}, compared to the performance at small-scale and large-scale, \method{} (w/ comm opt.) achieves a higher speedup over \method{} (w/o comm opt.) at medium-scale cases as communication is throughput-bound. In large-scale cases where communication becomes latency-bound, the speedup decreases. Nevertheless, the best performance of \method{} (w/ comm) still surpasses that of \method{} (w/o comm). These experimental results are consistent with the conclusion of our theoretical analysis presented in Sec.~\ref{sec:perf_gain} and Fig.~\ref{fig/method:perf_analysis}.

\begin{table*}
    \centering
    \caption{Performance and accuracy comparison of different GPU baselines on large-scale datasets. The results are for the highest achieved performance on the largest machine scale runs reported by the original papers. The size of datasets increases from left (Ogbn-products) to right (IGB260M).}
    \label{table/experiment:gpu_cpu_baselines}
    \resizebox{0.93\textwidth}{!}{%
    \begin{tabular}{|l|c|cc|cc|cc|cc|cc|}
        \hline
        \multirow{2}{*}{\textbf{Methods}} & \multirow{2}{*}{\textbf{Platform}} & \multicolumn{2}{c|}{\textbf{Ogbn-products}} & \multicolumn{2}{c|}{\textbf{Reddit}} & \multicolumn{2}{c|}{\textbf{Ogbn-papers100M}} & \multicolumn{2}{c|}{\textbf{Ogb-lsc-mag240M}} & \multicolumn{2}{c|}{\textbf{IGB260M}} \\ \cline{3-12}
        & & Time (s) & Acc. (\%) & Time (s) & Acc (\%) & Time (s) & Acc. (\%) & Time (s) & Acc. (\%) & Time (s) & Acc. (\%) \\
        \hline
        \textsf{DGL~\cite{zheng2020distdgl}} & GPU & 0.99 & 79.19  & 7.28 & 97.1  & 17.0 & --   & --  & --   & --  & --   \\
        \textsf{PipeCCN~\cite{wan2022pipegcn}} & GPU  & 0.43 & 78.77  & 0.43 & 97.10 & 6.70 & --   & --  & --   & --  & --   \\
        \textsf{BNSGCN~\cite{wan2022bns}} & GPU  & 0.28 & 79.30  & 0.19 & \textbf{97.15} & \textbf{0.59} & --   & --  & --   & --  & --   \\
        \textsf{AdapQ~\cite{wan2023adaptive}} & GPU  & 0.47 & 78.90  & 0.38 & 96.53 & --   & --   & --  & --   & --  & --   \\
        \textsf{SYLVIE~\cite{zhang2024sylvie}} & GPU  & 0.23 & 78.85  & 0.50 & 96.87 & 1.30 & --   & --  & --   & --  & --   \\
        \textsf{\textbf{SuperGCN (ours)}} & CPU  & \textbf{0.07} & \textbf{80.24}  & \textbf{0.13} & 96.55 & 0.65 & 65.63 & \textbf{0.80} & \textbf{65.29} & \textbf{1.62} & \textbf{54.93} \\
        \hline
    \end{tabular}
    }
    \\
\small
Time (s): Epoch time in seconds \quad Acc.: Test accuracy (Val accuracy for mag240M) \quad The results of DGL are collected from ~\cite{zhang2024sylvie}.
\end{table*}

\begin{figure*}[ht]
    \centering
    \includegraphics[width=0.97\linewidth]{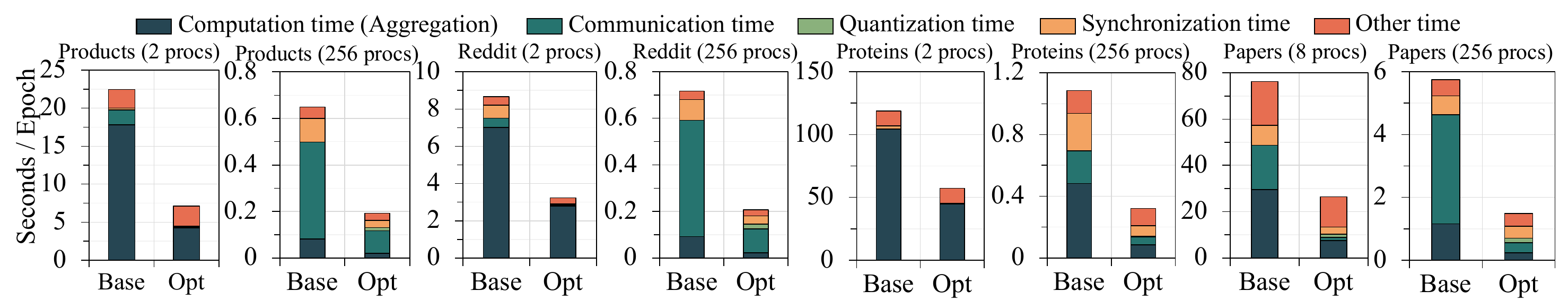}
    \caption{Time breakdown of different datasets on ABCI (Intel x86). \zhuang{\textit{Base} represents vanilla PyG implementation, while \textit{Opt} refers to SuperGCN with all our proposed optimizations.}}
    \label{fig/experiment:breakdown_abci}
\end{figure*}

\subsection{Accuracy Evaluations on Multiple CPUs}\label{sec:quan_acc_exp}

To assess the impact of our proposed communication-aware quantization scheme on model convergence and accuracy, we report the model convergence and accuracy on real large-scale datasets. 
We measure model convergence by monitoring changes in accuracy over training epochs, in all datasets. Additionally, we use the final test accuracy to report the model accuracy. For Ogb-lsc-mag240M, we use the validation accuracy, since this dataset has no test label. 
\zhuangppopp{The results can be divided into two parts: results of DistGNN, \method{} for Reddits and Ogbn-products on ABCI, and results of \method{} for Ogbn-papers100M, Ogb-lsc-mag240M, and IGB260M on Fugaku. For \method{}, }
We employ four settings to check the impact of Int2 quantization communication and masked label propagation on model accuracy: FP32 without masked label propagation (w/o LP), FP32 with masked label propagation (w/ LP), Int2 w/o LP, and Int2 w/ LP. We compared our final model accuracy after convergence with other baselines (shown in Table~\ref{table/experiment:gpu_cpu_baselines}) and the OGB leaderboard~\cite{ogb-leaderboard} across various datasets, demonstrating that our model accuracy is within a reasonable range.

Accuracy for different datasets is shown in Fig.~\ref{fig/experiment:accuracy_combo_abci}. \zhuangppopp{
In comparison to DistGNN, on Ogbn-products, even if our method converges at the same rate as DistGNN, our approach achieves higher accuracy on the test dataset after model convergence. On Reddit, there are some issues with the implementation of DistGNN, which result in abnormal training curves. 
On Ogbn-products, Reddit, and IGB260M, 
we found that Int2 closely matches FP32 regardless of whether label propagation is used. However, enabling masked label propagation allows the model to converge faster. On large-scale datasets Ogbn-papers100M and Ogb-mag240M, Int2 converges to lower accuracy than FP32 without masked label propagation, yet matches FP32 when label propagation is enabled. 
This empirically validates that the masked label propagation we introduce mitigates the accuracy drop caused by Int2 communication on large-scale datasets.}

Furthermore, Table~\ref{table/experiment:accuracy} presents a comparison of test accuracy between DistGNN and the different settings of \zhuang{\method{}} after model convergence. 
\zhuangppopp{
Similar to the result shown in Fig.~\ref{fig/experiment:accuracy_combo_abci}, masked label propagation does not impact model accuracy for Ogbn-products, Reddit, and IGB260M. However, on Ogbn-papers100M and Ogb-mag240M, masked label propagation reduces the accuracy loss by Int2. Empirical results demonstrate that our communication-aware algorithm design maintains consistent training convergence and model accuracy.}

\textbf{SuperGCN (full-batch) versus SoTA Mini-batch Training}
In Table~\ref{table/experiment:accuracy} we also report the highest reported mini-batch accuracy in literature using exactly the same model (GraphSAGE). \method{} consistently achieves higher (or similar) accuracy to SoTA mini-batch solutions. 

\subsection{Performance and Accuracy comparison with GPU baselines}
In this section, we compare our framework on supercomputer Fugaku with the current state-of-the-art GPU baseline in terms of performance and model accuracy across different datasets. We collect the performance (seconds per epoch) and model accuracy for the highest achieved performance on the largest machine scale runs reported by the original papers. The GPU baselines are optimized for distributed full-batch GCN training, including AdaptQ~\cite{wan2023adaptive} and SYLVIE~\cite{zhang2024sylvie} that employ adaptive quantization communication for improving framework's performance and scalability. We increase our epoch number to align with their setting for this comparison. The results are listed in Table~\ref{table/experiment:gpu_cpu_baselines}. We can summarize the results: (1) Scaling to the highest performance allowed before performance regresses: our method (\method{}) achieves the best performance on the ogbn-products and Reddit datasets while maintaining model accuracy within a normal range. (2) On the Ogbn-papers100M dataset, converting the directed graph to undirected improves accuracy but increases the computation and communication costs. Our approach achieves high accuracy using an undirected graph while maintaining near-best performance. Unlike with other datasets, our experiments (Table~\ref{table/experiment:accuracy}) show lossy communication reduces accuracy on this dataset. However, our method preserves model accuracy even when using lossy communication techniques. (3) Our approach also demonstrates performance and model accuracy on the mag240M and IGB260M datasets. To our knowledge, they are the largest publicly available GNN training datasets. All these results further demonstrate that, compared to state-of-the-art baselines, our framework achieves faster training while maintaining similar model accuracy and remains effective on large-scale datasets.

\subsection{Runtime Breakdown on Multiple CPUs}
To understand performance bottlenecks when using \method{} for full-batch GCNs training, we collect the breakdown of training time before and after using our proposed optimization method on both small-scale and large-scale runs (shown in Fig.~\ref{fig/experiment:breakdown_abci}). 
We divide the training process into 5 components: (a) \textit{Aggr time}, time of aggregation operation in GCN layers, (b) \textit{Comm time}, time of communication in GCN layers, (c) \textit{Quant time}, time of quantization and dequantization in GCN layers, (d) \textit{Sync time}, time of synchronization in GCN layers (check load imbalance), (e) \textit{Other time}, time spent on other components of training. 
To collect precise results, we switch off the overlapping of computation and communication. 

For small graphs, the performance bottleneck is the aggregation operation within the GCN layer for Ogbn-products, Reddit, and Proteins. Therefore, when using our proposed method designed for a single CPU, the time spent on the aggregation operation is significantly reduced, and the proportion of aggregation operation time in the total training time also decreases significantly. 
For large-scale runs, the performance bottleneck shifts to communication. After employing our proposed optimizations for reducing communication volume, the communication time drops significantly. 

\begin{table}
\centering
    \caption{\zhuang{Communication volume and time in 1 GCN layer under different communication methods. This test is conducted for Ogb-lsc-mag240M dataset on Fugaku (2048 procs).}}
    \resizebox{\linewidth}{!}{
        \begin{tabular}{|ll|c|c|}
            \hline
            \multicolumn{2}{|l|}{\bf{Methods}}                                                                                           & \multicolumn{1}{l|}{\bf{Comm volume (GB)}} & \multicolumn{1}{l|}{\bf{Comm time (ms)}} \\ \hline
            \multicolumn{2}{|l|}{\textsf{SuperGCN (pre\_aggr)}}                                                                                  & 1934.8559                                      & 1094.35                             \\ 
            \multicolumn{2}{|l|}{\textsf{SuperGCN (post\_aggr)}}                                                                                 & 1934.8559                             & 1131.62                             \\ 
            \multicolumn{2}{|l|}{\textsf{SuperGCN (pre\_post\_aggr)}}                                                                            & 1269.5784                            & 730.792                             \\ \hline
            \multicolumn{1}{|l|}{\multirow{2}{*}{\begin{tabular}[c]{@{}l@{}}\textsf{SuperGCN }\\ \textsf{(pre\_post\_aggr+Int2)}\end{tabular}}} & \textsf{data}   & 80.4770                               & 47.0393                             \\ 
            \multicolumn{1}{|l|}{}                                                                                  & \textsf{params} & 1.6530                                & 4.82566                             \\ 
            \hline
        \end{tabular}
    }
    \label{table/experiment:comm_statistics_fugaku}
\end{table}

\subsection{\zhuang{Efficacy of Communication Optimizations}}
In this section, we examine the effectiveness of our proposed optimizations on communication. Various configurations are evaluated, including (1) \emph{Pre}: solely applying pre-aggregation, (2) \emph{Post}: solely applying post-aggregation, (3) \emph{Pre-post}: applying the hybrid method of pre-aggr and post-aggr, (4) \emph{Pre-post+Int2}: combining \emph{Pre-post} with Int2 quantization. Here, \emph{data} refers to the communication of quantized feature vectors, while \emph{param} refers to the communication of parameters (zero point and scale) utilized for dequantization. 
Table~\ref{table/experiment:comm_statistics_fugaku} shows that our proposed hybrid of pre-aggr and post-aggr (pre-post-aggregation) method reduces communication volume and time by approximately 1.5$\times$ in comparison to \emph{Pre} or \emph{Post}. Despite the extra communication overhead introduced for dequantization, it still decreases the communication volume and time by about 15$\times$. These results demonstrate the efficacy of our proposed optimizations on communication.

\section{Related Work}
There are several GCN training frameworks optimized for mini-batch training, such as Aligraph~\cite{yang2019aligraph}, P3~\cite{gandhi2021p3}, DistDGL~\cite{zheng2020distdgl}, 
etc. However, a detailed study~\cite{jia2020roc} indicates sampling-based methods suffer from model accuracy loss. 

On GPU platforms, several optimization methods have been presented for aggregation operators~\cite{huang2020gespmm, fu2022tlpgnn, gale2020sparse}. On CPU platforms, DistGNN~\cite{md2021distgnn} optimizes aggregation operators (SpMM) for Intel CPUs using Intel LibXSMM~\cite{heinecke2016libxsmm}. However, it primarily targets Intel x86 CPUs. In addition, there are approaches proposed for both CPUs and GPUs~\cite{hu2020featgraph, ye2023sparsetir} based on DL compilers (e.g. TVM~\cite{chen2018tvm}).

We categorize the communication optimization methods into three main approaches: (1) some methods~\cite{md2021distgnn, wan2022pipegcn, peng2022sancus, thorpe2021dorylus} aim to mitigate the communication overhead by overlapping asynchronous communication with computation in the subsequent epoch. However, asynchronous communication introduces staleness for nodes' features, resulting in slower training convergence~\cite{dai2018toward}. (2) optimizing communication paths based on network topology. DGCL~\cite{cai2021dgcl} constructs a weighted graph to characterize the network topology and employs a tree-based algorithm to identify the optimal communication path. 
(3) reducing communication volume to mitigate the communication cost. CAGNET~\cite{tripathy2020reducing} employs a communication-avoiding algorithm to reduce communication volume. BNS-GCN~\cite{wan2022bns} lowers communication volume by randomly sampling the boundary nodes to transfer. However, this approach modifies the graph structure. AdapQ~\cite{wan2023adaptive},  SYLVIE~\cite{zhang2024sylvie} and \cite{zhuang2024communication} introduce stochastic integer quantization~\cite{chen2021actnn} to compress boundary nodes' features.
However, to obtain a trade-off between accuracy and performance, they have extra overhead to select an appropriate combination of 2, 4, and 8 bits for quantization, whereas our method applies a uniform 2-bit quantization communication, hereby reducing the communication volume to minimum, while also avoiding the high cost of adaptive quantization.



\section{Conclusion}
\label{sec:conclusion}

\zhuanghpdc{We present \method{}, a distributed training framework for graph convolutional networks designed specifically for CPU-based supercomputing systems. Our framework addresses the critical challenge of irregular memory access and communication overhead in distributed full-batch GCN training through three major contributions: general and efficient aggregation operators, a hierarchical aggregation scheme that reduces communication costs while preserving graph structure, and a communication-aware quantization scheme that effectively utilizes quantized communication while maintaining model accuracy. \method{} achieves a speedup of up to 6$\times$ in comparison to SoTA CPU-based implementations, scales to thousands of processors on the largest publicly available datasets, and outperforms SoTA GPU-based distributed full-batch GCN training frameworks at peak performance.}


\begin{acks}
The authors wish to express their sincere gratitude to Tal Ben-Nun from Lawrence Livermore National Laboratory for his invaluable advice throughout this work.
\end{acks}

\bibliography{0.ref}




\end{document}